\documentclass[sigconf]{acmart}
\AtBeginDocument{%
  }

\setcopyright{acmlicensed}
\copyrightyear{2026}
\acmYear{2026}
\acmDOI{XXX}
\acmConference[SIGMOD]{SIGMOD}
\acmISBN{}

\usepackage{xcolor}
\usepackage{etoolbox}
\usepackage{booktabs}
\usepackage{array,longtable,url}
\usepackage[normalem]{ulem}

\begin{document}

\title{SIGMA: A Versatile Streaming Graph Partitioner for Vertex- and Edge-Balanced Distributed GNN Training}

\author{Barbara Hoffmann}
\authornotemark[1]
\email{barbara.hoffmann@uni-bayreuth.de}
\orcid{0009-0004-2963-1882}
\affiliation{%
  \institution{University of Bayreuth}
  \city{Bayreuth}
  \country{Germany}
}

\author{Shai Dorian Peretz}
\authornote{Both authors contributed equally to this research.}
\email{uowky@student.kit.edu}
\affiliation{%
  \institution{Karlsruhe Institute of Technology}
  \city{Karlsruhe}
  \country{Germany}
}

\author{Adil Chhabra}
\email{adil.chhabra@informatik.uni-heidelberg.de}
\orcid{0009-0009-5726-9389}
\affiliation{%
  \institution{Heidelberg University}
  \city{Heidelberg}
  \country{Germany}
}

\author{Ahmet Kadir Yalcinkaya}
\email{ahmet.yalcinkaya@uni-bayreuth.de}
\affiliation{%
  \institution{University of Bayreuth}
  \city{Bayreuth}
  \country{Germany}
}

\author{Ruben Mayer}
\email{ruben.mayer@uni-bayreuth.de}
\orcid{0000-0001-9870-7466}
\affiliation{%
  \institution{University of Bayreuth}
  \city{Bayreuth}
  \country{Germany}
}

\author{Christian Schulz}
\email{christian.schulz@informatik.uni-heidelberg.de}
\orcid{0000-0002-2823-3506}
\affiliation{%
  \institution{Heidelberg University}
  \city{Heidelberg}
  \country{Germany}
}



\begin{abstract}
Distributed Graph Neural Network (GNN) training depends critically on how the underlying graph is partitioned across compute resources.
Existing graph partitioners focus either on vertex partitioning or edge partitioning and typically optimize only a single communication objective (edge cut \emph{or} vertex cut) under a single balance constraint (vertex balance \emph{or} edge balance). 
We present SIGMA (Streaming Integrated Graph Partitioning with Multi-objective Awareness), a versatile streaming graph partitioner that supports both vertex and edge partitioning within a unified multi-objective, multi-constraint framework. Depending on the target distributed GNN system, SIGMA can be configured for edge-cut-oriented vertex partitioning or vertex-cut-oriented edge partitioning while simultaneously accounting for \emph{both} vertex and edge balancing. A clustering-based preprocessing stage incorporates global graph structure to improve partition quality while preserving the efficiency and scalability advantages of streaming partitioning. We evaluate SIGMA on six benchmark graphs spanning diverse domains and scales using two distributed GNN training systems: DistGNN (edge-partitioned) and DistDGL (vertex-partitioned). Across both settings, SIGMA consistently achieves strong performance, showing its ability to navigate complex trade-offs between partition quality, training efficiency, and memory consumption, 
frequently outperforming established streaming approaches and delivering competitive results against high-quality in-memory partitioners, with particularly strong results in edge-partitioned GNN training. 
These results demonstrate that a unified streaming partitioner can effectively address the communication, compute, and memory challenges of distributed GNN training across fundamentally different system architectures.
\end{abstract}



\maketitle



\section{Introduction}
Graph Neural Networks (GNNs) have become fundamental for learning over large-scale graph-structured data, with applications in social networks~\cite{shrivastava2025socialnetworks}, drug discovery~\cite{xiong2021drugdiscovery, li2021drugdiscovery}, fraud detection~\cite{liu2021frauddetection, shi2022h2frauddetection}, natural language processing~\cite{wu2023NLP}, and recommender systems~\cite{wu2022recommenderSystems}. As graphs grow to millions or billions of vertices and edges~\cite{hu2020opengraphbenchmark}, distributed training becomes essential, with its efficiency strongly depending on graph partitioning across compute resources~\cite{merkel2023experimental, 10.14778/3648160.3648167}.

Graph partitioning affects computational balance and communication overhead in GNN training~\cite{merkel2023experimental}, as observed in other graph analytics and graph databases~\cite{10.14778/3236187.3236208, 10184652}. Uneven distributions of training vertices can also slow model convergence~\cite{10.14778/3648160.3648167}. However, extensive experimental studies show that traditional partitioners are ill-suited to these requirements~\cite{merkel2023experimental, 10.14778/3648160.3648167}. Conventional vertex partitioning assigns vertices to blocks and cuts edges, whereas edge partitioning assigns edges to blocks and cuts vertices. Both formulations balance only one element type rather than vertices and edges simultaneously, despite GNN training requiring both vertex-centric aggregation and edge-centric message passing. Moreover, the literature largely treats vertex and edge partitioning separately, duplicating research efforts and slowing progress.

To address these shortcomings, we introduce \textsf{SIGMA}, a streaming graph partitioner that unifies vertex and edge partitioning, supports independent or joint optimization of edge cut and replication factor, and simultaneously enforces vertex and edge balance. We evaluate partitioning characteristics and end-to-end distributed GNN training performance, demonstrating that improved partitioning translates into more efficient training.

The main contributions of this work are:
\begin{itemize}
    \item We present \textsf{SIGMA}, a multi-objective, multi-constraint streaming graph partitioner that optimizes edge cut, replication factor, or both while enforcing simultaneous vertex and edge balance under explicit capacity constraints. This is challenging because objectives and constraints are coupled: balancing vertices can induce edge imbalance and vice versa. \textsf{SIGMA} addresses these trade-offs while accounting for the communication and memory requirements of distributed GNN training, enabling more efficient and scalable model training.

    \item We introduce a unified streaming partitioning framework that supports both vertex and edge partitioning within a single approach, bridging the gap between two traditionally separate lines of graph partitioning research.

    \item We release an efficient open-source implementation of \textsf{SIGMA} at \url{https://github.com/code-SIGMA/SIGMA}.

    \item We conduct a comprehensive evaluation on six benchmark graphs using DistGNN (edge-partitioned, full-batch) and DistDGL (vertex-partitioned, mini-batch). We assess partitioning quality (replication factor or edge cut ratio), vertex and edge balance, partitioning time, 
    and training efficiency in terms of epoch time and memory consumption.
\end{itemize}

\section{Background} \label{sec:2_background}
This section provides the necessary background on GNNs and graph partitioning, with a focus on their interaction in distributed \hbox{training settings.}

\subsection{Graph Partitioning} \label{subsec:2.1_graphPartitioning}

\subsubsection{Problem Formalization}

We consider an undirected graph $G = (V , E)$ with no parallel or self-edges with $n = |V|$ vertices and $m = |E|$ edges. Let $c: V \rightarrow \mathbb{R}_{\ge 0}$ denote the vertex weight function and $\omega : E \rightarrow \mathbb{R}_{> 0}$ the edge weight function. The weight functions $c$ and $\omega$ are generalized additively to sets. An edge $e = (u,v)$ is said to be incident on vertices $u$ and $v$. The set $N(v) = \{ u \in  V \mid \ (u,v) \in E\}$ describes the neighborhood of $v$, and the degree as $d(v) = |N(v)|$. Given an integer $k \ge 1$ the goal of graph partitioning is to divide the graph into $k$ disjoint blocks while optimizing an objective function under balance constraints. In literature, two ways of partitioning a graph have been considered: vertex partitioning and edge partitioning.

A $k$-way \textbf{vertex partitioning} of $G$ divides the vertex set $V$ into $k$ disjoint blocks $V_1, \dots, V_k$ with $V_i \cap V_j = \emptyset \text{ for } i \ne j$ such that $\bigcup_{i = 1}^k V_i = V$. The optimization objective is to minimize the \emph{edge-cut} $\omega(E_{\text{cut}})$, where $E_{\text{cut}} := \{ (u,v) \in E \mid u \in V_i, v \in V_j, i \ne j\}$, denoting the total weight of edges whose endpoints lie in different blocks, subject to the vertex balance constraint $c(V_i) \le L_{\text{max}} := \lceil (1+\epsilon) \frac{c(V)}{k} \rceil$ for a given imbalance parameter $\epsilon \ge 0$. The minimization of the edge-cut results in a lower inter block communication, as less inter-vertex messages need to be transferred across blocks.

A $k$-way \textbf{edge partitioning} of $G$ divides the edge set $E$ into $k$ disjoint blocks $E_1, \dots, E_k$, such that $\bigcup_{i = 1}^k E_i = E$. The optimization objective is to minimize the \emph{replication factor}, which measures the average number of partitions each vertex is replicated across. Let the set \hbox{$V(E_i) = \{ v \in V \mid \exists u \in V: (u,v) \in E_i\}$} and $|V(E_i)|$ describe the number of vertices that have at least one incident edge to a vertex in $E_i$. More formally, we seek to minimize the replication factor defined as $RF(E_1, E_2, \dots, E_k) = \frac{1}{n} \sum_{i = 1}^k |V(E_i)|$, subject to the edge balance constraint \hbox{$\omega(E_i) \le L_{\text{max}} := \lceil (1 + \epsilon) \frac{\omega(E)}{k}\rceil $}. Minimizing the replication factor reduces synchronization overhead in distributed graph processing by limiting the need to synchronize vertex state between partitions.

\subsubsection{Partitioning Paradigms}
Graph partitioning methods can be broadly categorized into in-memory and streaming approaches. In-memory algorithms assume full random access to the input graph and typically perform multiple passes to iteratively refine partition assignments, yielding high-quality partitions at the cost of substantial memory consumption. In contrast, streaming partitioning algorithms process the graph incrementally, assigning vertices or edges online as they arrive. These methods follow a load–compute–store paradigm~\cite{ccatalyurek2023more}, assigning elements via a scoring function derived from local neighborhood information and limited partition state.


Streaming partitioning approaches can be further divided into \emph{vertex-streaming} and \emph{edge-streaming} models. In vertex-streaming partitioning, the complete adjacency list (neighborhood) of a vertex is assumed to be available when the vertex arrives and is assigned to a block. In contrast, edge-streaming partitioning exposes the graph as a stream of individual edges, each of which must be assigned upon arrival. 
Our framework supports both paradigms, enabling either streamed assignment of vertices to blocks or streamed assignment of edges to blocks.
Streaming approaches can additionally be categorized as \emph{stateless} or \emph{stateful}. Stateless methods assign elements independently of prior assignments, typically using simple heuristics such as hashing, which leads to poor partition quality due to the lack of structural awareness. Conversely, stateful methods maintain lightweight graph properties, such as block assignment and/or replica information, to guide partitioning decisions.

\paragraph{Streaming Model.}
Throughout this paper, streaming refers to the \emph{access model}, not to the input: we partition a static, finite graph that exists in full, e.g., on disk, but is processed sequentially rather than loaded into memory. This is distinct from partitioning \emph{streaming or dynamic graphs}, where the graph itself changes over time. The input is consumed as a finite sequence of elements, i.e., vertices together with their adjacency lists or individual edges, arriving in an arbitrary but fixed order. Each element must be assigned to a block immediately and irrevocably upon arrival; its associated information, such as the adjacency list, is then discarded, and only compact partition state, such as block loads and vertex-to-block assignments, is retained across the stream. Algorithms in this model may perform a small, constant number of such sequential passes over the input: one-pass methods such as \textsf{Fennel} and \textsf{HDRF} read the graph once, whereas multi-pass methods such as \textsf{2PS}, and \textsf{SIGMA} with its clustering, preassignment, and assignment passes (Section~\ref{sec:approach_partitioning}), use additional passes to incorporate global structure while preserving the same per-pass access model.



\paragraph{Representative Streaming Algorithms.}
A representative stateful vertex-streaming partitioning framework is \textsf{Fennel}~\cite{tsourakakis2014fennel}, which assigns each arriving vertex $v$ to the block $p$ that maximizes
\[
p^* = \arg\max_p \; C_{\text{\textsf{FENNEL}}}(v,p) = |N(v) \cap V_p| - \alpha \cdot \gamma \cdot V_p^{\gamma - 1},
\]
where $V_p$ is the set of vertices already assigned to block $p$, and $\alpha,\gamma > 0$ are parameters controlling the trade-off between edge-cut minimization and vertex balance.

Similarly, in stateful edge-streaming partitioning, an edge $e = (u,v)$ is assigned to the block $p$ that maximizes a scoring function $C(u,v,p)$. A prominent example is \textsf{HDRF}~\cite{petroni2015hdrf}, which selects
\[
p^* = \arg\max_p \; C_{\text{\textsf{HDRF}}}(u,v,p) = C_{\text{REP}}(u,v,p) + C_{\text{BAL}}(p),
\]
where $C_{\text{REP}}(u,v,p)$ is a degree-weighted replication score that favors blocks already containing $u$ and/or $v$, and $C_{\text{BAL}}(p)$ is a balancing term that favors blocks with fewer assigned edges. This formulation encourages co-location of adjacent vertices while maintaining balanced edge distributions.

\subsection{Distributed Graph Neural Network Training} \label{subsec:2.2_GNNTraining}
\subsubsection{Graph Neural Networks}

Modern GNNs learn vertex representations through iterative
message passing along graph edges, aggregating and transforming
information from neighboring vertices~\cite{kipf2016semi,hamilton2017inductive}.
Let $G=(V,E,X)$ denote a graph with vertex set $V$, edge
set $E$, and feature matrix $X\in\mathbb{R}^{|V|\times d}$
containing $d$-dimensional vertex features.
For each $v\in V$, let $\mathcal{N}(v)$ denote its
neighborhood and $h_v^{(k)}$ its representation at layer $k$,
initialized with its feature vector $x_v$:
$h_v^{(0)}=x_v$.

Each message-passing layer consists of aggregation and update.
First, vertex $v$ aggregates neighbor representations:
\begin{equation}
m_v^{(k+1)} = \mathrm{AGGREGATE}^{(k)}\left(
\left\{ h_u^{(k)} \mid u \in \mathcal{N}(v) \right\}\right),
\end{equation}
where $\mathrm{AGGREGATE}^{(k)}$ is permutation-invariant
(e.g., sum, mean, max, or attention-weighted sum).
The aggregated message is then used to update the vertex
representation:
\begin{equation}
h_v^{(k+1)} = \mathrm{UPDATE}^{(k)}\left(
h_v^{(k)}, m_v^{(k+1)}\right).
\end{equation}
The update function typically applies a learnable
transformation (e.g., a linear layer or multi-layer
perceptron) followed by a nonlinear activation.

Thus, vertex representations depend on both their own
features and their neighbors' representations, tightly
coupling computation to the graph structure.
Large-scale GNN training is therefore typically distributed
across multiple GPUs on multiple servers~\cite{zheng2020distdgl}.

GNN training is commonly performed using either full-graph training or mini-batch sampling techniques. In full-graph training, the entire graph is processed in each training iteration, which provides exact neighborhood information but requires the full graph to reside in memory and often leads to substantial communication overhead. In contrast, mini-batch approaches sample subgraphs or neighborhoods for each iteration, reducing memory requirements and increasing scalability.


\subsubsection{Distributed Training}
\label{subsec:requirements_gnn}

Distributed GNN training involves executing the aggregation and update steps across a number of compute workers; these could be assigned to compute nodes or devices (such as GPUs) on a node. To achieve distributed training, the system assigns ownership of graph data to specific workers according to the underlying partitioning paradigm. 

\paragraph{Vertex Partitioning}

\begin{figure}
    \centering
    \includegraphics[width=0.8\linewidth]{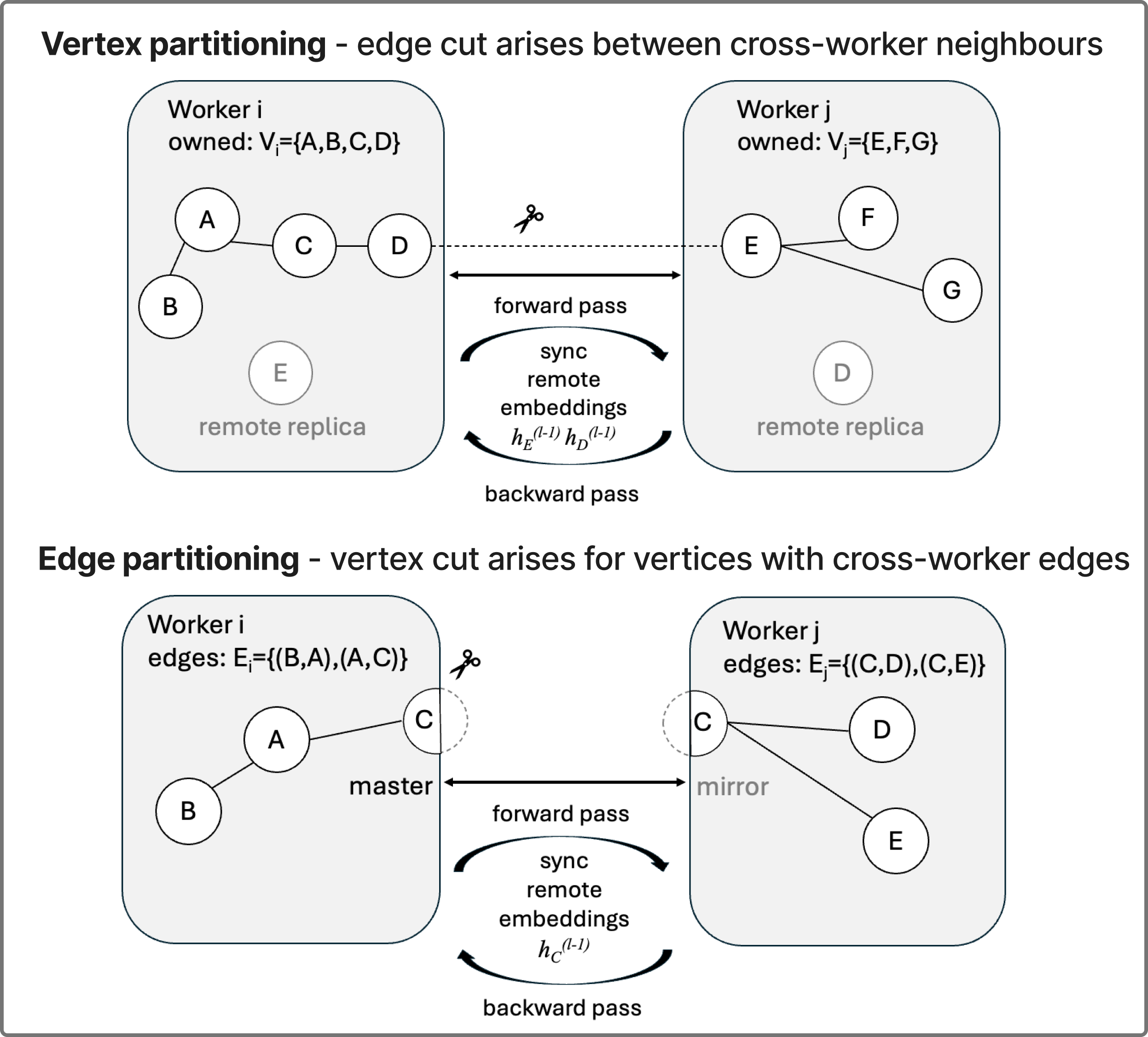}
    \caption{Comparison of GNN partitioning paradigms for distributed training.}
    \label{fig:gnnParadigms}
\end{figure}

Worker $i$ typically owns a set of vertices $V_i$ and stores their $d$-dimensional feature tensors $x_v$, intermediate embeddings $h_v^{(l)}$, and optimizer states. Aggregation is the primary synchronization point: each iteration (i) synchronizes remote embeddings, (ii) performs local aggregation and updates, and (iii) maintains consistency of replicated vertex states across workers. Computing $h_v^{(l)}$ for $v \in V_i$ requires the $(l-1)$-level embeddings of all neighbors $u \in \mathcal{N}(v)$.
If a neighbor $u$ resides on a different worker $j$, the system must perform a network transfer to synchronize the state, typically using \emph{remote replicas} (or ghost vertices) on worker $i$ that are updated during each forward and backward pass as shown in Figure~\ref{fig:gnnParadigms}. Ghost vertices induce additional memory consumption, as a copy of their state needs to be held.

Thus, the partitioning strategy directly dictates the lifecycle of data movement and the memory footprint of vertex replicas. Under this model, an \emph{edge cut} occurs whenever two connected vertices are assigned to different workers.

\paragraph{Edge Partitioning}
Here, the graph is partitioned by distributing subsets of edges $E_i$ across workers. As a consequence, vertices may appear in multiple partitions whenever their connecting edges are assigned to different workers.
During message passing, worker $i$ processes locally assigned edges $(u,v) \in E_i$ but still requires consistent embeddings for all participating vertices. If the edges connected to a vertex $v$ are distributed across multiple workers, the corresponding embedding $h_v^{(l)}$ must be synchronized across partitions during forward and backward propagation as shown in Figure \ref{fig:gnnParadigms}.
Under this model, a \emph{vertex cut} occurs when the edges connected to a vertex are distributed across multiple workers.
Distributed systems therefore typically maintain one primary copy (master vertex) together with replicated mirror copies on remote workers to synchronize shared vertex states across partitions.
Consequently, communication arises from synchronizing replicated vertex states across workers, while memory overhead is determined by the resulting replication factor. 

\paragraph{System-Level Objectives for Distributed GNN Partitioning.}
Distributed GNN training requires the simultaneous optimization of three objectives to address specific hardware bottlenecks. While these objectives are presented separately, they are interdependent in practice:

 \textbf{Vertex Balance:} Vertex-associated data such as vertex features can substantially affect memory consumption in distributed GNN training systems~\cite{merkel2023experimental}. In distributed settings, each worker is constrained by its local memory capacity, whether GPU VRAM or CPU main memory. Imbalanced vertex assignments can, therefore, lead to out-of-memory failures on individual workers, even when aggregate system memory is sufficient. In addition, memory usage is further affected by factors such as intermediate activations and vertex replication, making balanced vertex partitioning a primary requirement for scalability.
 
  \textbf{Edge Balance:} The computational workload of message passing is largely driven by the number of processed edges, making edge count a useful proxy for aggregation cost. However, total computation is also influenced by feature dimensionality and vertex-wise update functions. In many distributed GNN systems, training proceeds in a (semi-)synchronous manner, requiring workers to synchronize at each layer. Edge imbalance can therefore create straggler workers that delay global progress, reducing hardware utilization and increasing overall training time.

 \textbf{Communication:} Communication arises whenever embeddings of remote neighbors must be exchanged across workers during message passing. The exact communication pattern depends on the partitioning strategy: in vertex partitioning, communication is largely driven by the number of cut edges, whereas in edge partitioning it is governed by the degree of vertex replication. In both cases, communication volume scales with embedding dimensionality, making data movement a major bottleneck on both network and memory subsystems. Furthermore, replication increases not only communication but also memory consumption, reinforcing the coupling between objectives~\cite{merkel2023experimental}.

These three objectives are inherently interdependent. For instance, enforcing strict vertex balance to satisfy memory constraints may necessitate a higher edge-cut, thereby increasing communication overhead. As feature dimensionality, model depth, and graph size increase, these overheads become more important, amplifying the importance of high-quality partitioning. Effective partitioning for GNNs must navigate these trade-offs to ensure that no single resource  (memory, compute, or network) becomes a bottleneck that limits system scalability.

\section{\textsf{SIGMA} Algorithm}
\label{sec:approach_partitioning}

As Section~\ref{subsec:requirements_gnn} shows, partitioning for distributed GNN training must simultaneously control communication (vertex replication and edge cut), computation (edge balance), and memory (vertex and replica balance). Existing partitioning approaches typically optimize edge cut with vertex balance or replication factor with edge balance, capturing only subsets of these objectives. We therefore need a partitioner that jointly optimizes edge cut and vertex replication while enforcing vertex and edge balance simultaneously. In addition, the partitioner should be fast, resource-efficient, and able to \hbox{handle large graphs.}


We therefore consider a streaming framework that supports multiple partitioning objectives within a unified setting. Our partitioner \textsf{SIGMA} (\underline{S}treaming \underline{I}ntegrated \underline{G}raph Partitioning with \underline{M}ulti-objective \underline{A}wareness) supports both vertex-streaming and edge-streaming partitioning: it can assign vertices to minimize edge cut, assign edges to minimize vertex replication, or jointly optimize both objectives within a single run. The partitioning mode itself is selected upfront to match the target GNN system (vertex- or edge-partitioned); joint optimization then takes place \emph{within} the chosen mode: the vertex-streaming score couples edge-cut locality with a replication penalty (Section~\ref{subsec:vertex_partitioning}), while the edge-streaming score couples replication affinity with edge- and replica-balance terms (Section~\ref{subsec:edge_partitioning}). In all cases, assignments are made online, using only the structural information revealed so far, and must satisfy multiple balance constraints across blocks.


When a stream element $x$ arrives, where $x$ is either a vertex or an edge depending on the partitioning mode, \textsf{SIGMA} assigns it to the feasible block that maximizes an assignment score:
$$
p^* = \arg\max_{p \in \mathcal{F}(x)} S(x,p),
$$
where $\mathcal{F}(x)$ denotes the set of feasible blocks for $x$. The score has the generic form
$$
S(x,p) = \operatorname{affinity}(x,p) - \operatorname{penalty}(x,p).
$$
The affinity term rewards locality, such as placing adjacent vertices or incident edges together, while the penalty term discourages imbalance and, when applicable, additional vertex replication.
This formulation is an \emph{a-priori} weighted-sum scalarization of multiple partitioning criteria: competing objectives are combined into a single score with fixed weights, rather than computing a Pareto front of trade-off solutions. It differs from a plain weighted sum in that scoring is restricted to the feasible set $\mathcal{F}(x)$ induced by multiple hard balance constraints, which are tightened dynamically over the stream as described below, and in that the same formulation instantiates both vertex- and edge-streaming partitioning.


The feasible set $\mathcal{F}(x)$ under multiple balance constraints is defined using a load vector $L_p$ and a capacity vector $U_p$, which \textsf{SIGMA} maintains for each block $p$. The dimensions of these vectors represent quantities controlled during partitioning, including assigned vertices, edge volume, edge load, and, when applicable, vertex replication. These dimensions map directly to the system-level bottlenecks of Section~\ref{subsec:requirements_gnn}: vertex and replica loads govern per-worker memory and thus the risk of out-of-memory failures, edge loads govern aggregation work and straggler-induced synchronization delays, and the scoring objectives target the communication volume induced by cut edges and vertex replicas. We write $L_{p,i}$ and $U_{p,i}$ for the load and capacity of block $p$ in dimension $i$, respectively. Assigning $x$ to block $p$ induces a load change $\Delta_x(p)$, whose $i$-th component is denoted by $\Delta_{x,i}(p)$. A block $p$ is feasible for $x$ if, for every hard-constrained balance dimension $i$,
$$
L_{p,i} + \Delta_{x,i}(p)
\leq
U_{p,i} \cdot \sigma(t).
$$

Here, $\sigma(t) = \sigma_{\min} + (1-\sigma_{\min})\sqrt{t}$ is a dynamic capacity scale and $t \in [0,1]$ denotes the fraction of elements streamed so far. The scale restricts admissible capacity early in the stream and gradually relaxes it to the full capacity bound $U_{p,i}$. This is important in the multi-constraint setting: if blocks are allowed to fill too quickly in one load dimension, later assignments may have few feasible choices, which can harm balance or partition quality. We set $\sigma_{\min}$ to the maximum relative block load after preprocessing defined in Section~\ref{subsec:preprocessing}, floored at $0.9$, i.e., 
$
\sigma_{\min}
=
\max\left\{
0.9,\;
\max_{p,i}
\frac{L_{p,i}}{U_{p,i}}
\right\}.
$
If no block is feasible, \textsf{SIGMA} assigns $x$ to the block that minimizes the maximum relative load after assignment.

The following subsections detail this framework for vertex and edge partitioning by specifying the corresponding loads, constraints, and scoring functions. For clarity, we present the formulation for unweighted graphs. Weighted edges can be handled by replacing edge counts and degrees with total edge weight and weighted \hbox{degrees, respectively.}

\subsection{Vertex Partitioning}
\label{subsec:vertex_partitioning}

In vertex partitioning, the stream element is a vertex $v \in V$, which must be assigned to exactly one block, denoted by $\pi(v)$. For each block $p$, \textsf{SIGMA} maintains the load vector
$$
L_p = \left(L_p^{\mathrm{vertex}}, L_p^{\mathrm{vol}}\right),
$$
where $L_p^{\mathrm{vertex}}$ counts assigned vertices and $L_p^{\mathrm{vol}}$ captures edge volume. Assigning a vertex $v$ to block $p$ contributes
$$
\Delta_v(p) = (1,d(v)+1).
$$

The first dimension enforces vertex balance, which, as discussed in Section~\ref{subsec:requirements_gnn}, directly impacts memory usage in distributed GNN training. The second dimension charges vertices proportionally to their degree $d(v)$, capturing their contribution to aggregation.  Summing $\Delta_v(p)$'s second component over all $v \in V$ gives $\sum_{v \in V}(d(v)+1) = 2|E|+|V|$ by the handshake lemma, the total volume to be balanced across the $k$ blocks. To ensure that both vertex and edge load remains balanced across blocks, we define the corresponding upper bounds for capacity:

$$
U_p^{\mathrm{vertex}} = \left\lceil (1+\varepsilon)\frac{|V|}{k}\right\rceil,
\qquad
U_p^{\mathrm{vol}} = \left\lceil (1+\varepsilon_E)\frac{2|E|+|V|}{k}\right\rceil.
$$

For an arriving vertex $v$, the locality of assigning $v$ to block $p$ is measured by
$$
e(v,p) = |\{u \in N(v) : \pi(u)=p\}|,
$$
i.e., the total number of edges from $v$ to already assigned neighbors in $p$. This yields a normalized variant of the classical \textsf{Fennel}~\cite{tsourakakis2014fennel} scoring function:
$$
S_{\mathrm{vertex}}(v,p) = \frac{e(v,p)}{d(v)} - \rho_p^{\gamma-1.1},
$$
where $\gamma > 1$ is a tunable parameter controlling the strength of the imbalance penalty, and
$$
\rho_p = \max\left\{
\frac{L_p^{\mathrm{vertex}}}{U_p^{\mathrm{vertex}}},
\frac{L_p^{\mathrm{vol}}}{U_p^{\mathrm{vol}}}
\right\}.
$$
The normalization by $d(v)$ ensures that the locality contribution is scale-invariant across vertices of different degrees, preventing high-degree vertices from dominating the score purely due to their larger neighborhoods. This follows the general structure of \textsf{Fennel}, where locality is balanced against an imbalance penalty, but extends it to a multi-dimensional load setting.

We further extend the vertex partitioning score with a replication-aware term. For a candidate assignment of $v$ to block $p$, let
$$
R(v,p) = R_1(v,p) + R_2(v,p),
$$
where $R_1(v,p)$ counts neighbors of $v$ that are not already present in $p$, and $R_2(v,p)$ counts the number of distinct neighbor blocks into which $v$ would additionally need to be replicated. More precisely, $R_2(v,p)$ counts the number of distinct blocks $\pi(u)$ of neighbors $u \in N(v)$ such that $\pi(u) \neq p$ and $v$ is not already present in $\pi(u)$.

The resulting multi-objective score is 
$$
S_{\mathrm{MO}}(v,p) = \frac{e(v,p)}{d(v)} - \rho_p^{\gamma-1.1} - \tau \cdot \frac{R(v,p)}{d(v)+k}.
$$
where $\tau > 0$ controls the strength of the replication penalty. The normalization by $d(v)+k$ scales the replication estimate relative to both the local neighborhood size and the number of blocks, ensuring that the penalty remains comparable across vertices of different degrees and across different values of $k$.
This additional term targets mirror vertex replication (ghost vertices, cf. Section~\ref{subsec:requirements_gnn}). It penalizes assignments likely to introduce additional vertex-to-block incidences, aligning local decisions with the global goal of minimizing vertex replication. Based on initial experiments, we propose the defaults $\gamma = 2.5$ and $\tau = 0.5$, which we found to provide a robust balance-quality trade-off. Increasing $\gamma$ weakens the imbalance penalty, prioritizing locality at the potential cost of load balance; increasing $\tau$ places greater weight on reducing vertex replication relative to locality.



\subsection{Edge Partitioning}
\label{subsec:edge_partitioning}

We additionally consider edge partitioning to optimize vertex replication directly. While the multi-objective vertex partitioner only estimates replication locally, edge partitioning controls it by colocating edges that share endpoints whenever possible, allowing the partitioner to reason explicitly about vertex-to-block incidences.


In edge partitioning, the stream element is an edge $(u,v) \in E$ which must be assigned to exactly one block, while vertices may be replicated in multiple blocks. Let $R_p \subseteq V$ denote the set of vertices currently replicated in block $p$, i.e., vertices incident to at least one edge assigned to $p$. The maintained load vector is
$$
L_p = \left(L_p^{\mathrm{rep}}, L_p^{\mathrm{edge}}\right),
$$
where $L_p^{\mathrm{rep}}$ counts vertex-to-block incidences (i.e., replicas) and $L_p^{\mathrm{edge}}$ counts the number of edges assigned to block $p$. Assigning an edge $(u,v)$ to block $p$ induces the load change
$$
\Delta_{(u,v)}(p)
=
\left(
\mathbf{1}[u \notin R_p] + \mathbf{1}[v \notin R_p],
\ 1
\right).
$$
The first dimension tracks newly created vertex replicas, while the second controls edge load. To ensure balance, the maximum capacity constraint for edge load is given by
$$
U_p^{\mathrm{edge}} = \left\lceil (1+\varepsilon_E)\frac{|E|}{k}\right\rceil.
$$
In contrast, replica load is not subject to a hard capacity constraint, as the number of vertex replicas depends on global assignments that cannot be bounded a priori in a streaming setting. Instead, replica load is incorporated into the scoring function to discourage uneven accumulation of replicas across blocks.

The assignment score follows the structure of \textsf{HDRF}~\cite{petroni2015hdrf}. Let $d(u)$ and $d(v)$ denote the degrees of the endpoints and define $s = d(u) + d(v)$. For a candidate block $p$, the endpoint-affinity terms are given by
$$
g_u(p)=
\begin{cases}
2 - \frac{d(u)}{s} & \text{if } u \in R_p,\\[6pt]
0 & \text{otherwise,}
\end{cases}
\qquad
g_v(p)=
\begin{cases}
2 - \frac{d(v)}{s} & \text{if } v \in R_p,\\[6pt]
0 & \text{otherwise.}
\end{cases}
$$
Thus, an endpoint contributes to the score only if it is already present in the block, encouraging assignments that reuse existing replicas. The contribution is larger for lower-degree vertices, introducing a bias toward preserving their placement.

For every feasible block $p$, \textsf{SIGMA} evaluates the score
$$
S_{\mathrm{edge}}(u,v,p)
=
g_u(p) + g_v(p)
+
\lambda \left(
\tfrac{1}{2} b_p^{\mathrm{edge}} + \tfrac{1}{2} b_p^{\mathrm{rep}}
\right),
$$
where $\lambda > 0$ controls the relative importance of balancing, and the balance terms are defined as
$$
b_p^{\mathrm{edge}} =
\frac{L_{\max}^{\mathrm{edge}} - L_p^{\mathrm{edge}}}
{\epsilon + L_{\max}^{\mathrm{edge}} - 1},
\qquad
b_p^{\mathrm{rep}} =
\frac{L_{\max}^{\mathrm{rep}} - L_p^{\mathrm{rep}}}
{\epsilon + L_{\max}^{\mathrm{rep}} - 1}.
$$

The term $b_p^{\mathrm{edge}}$ promotes balanced edge assignments across blocks, while $b_p^{\mathrm{rep}}$ penalizes blocks that accumulate a large number of vertex replicas. This extends classical \textsf{HDRF}, which balances edge load while minimizing replication, by additionally discouraging uneven distribution of vertex replicas. As discussed in Section~\ref{subsec:requirements_gnn}, vertex distribution directly impacts memory usage and training efficiency in GNN workloads. Penalizing blocks with high replica load indirectly regularizes replica distribution, mitigating memory and communication skew without explicit vertex balance.


\subsection{Clustering-Based Preprocessing}
\label{subsec:preprocessing}
The streaming partitioners described above make assignment decisions based only on the currently observed vertex or edge and some limited state about past assignments. As a result, they do not have access to global structural information, which can lead to suboptimal assignments when densely connected regions are fragmented across blocks. Many real-world graphs, however, exhibit pronounced community structure, where vertices within the same cluster share a large number of edges. Preserving such structure during partitioning reduces edge cut or vertex replication. 

The \textsf{2PS} algorithm~\cite{mayer20202ps} addresses this limitation for streaming edge partitioning by incorporating global structural information through a clustering preprocessing stage. It first computes a streaming clustering of the graph, then maps the resulting clusters to blocks using a makespan scheduling formulation, and finally uses the cluster-to-block mapping to prepartition edges whose endpoints are consistent w.r.t. cluster assignment. Edges that cannot be prepartitioned are subsequently handled by a streaming edge partitioner.

Our framework adopts this strategy, but generalizes it in three ways. First, while \textsf{2PS} is designed for streaming edge partitioning, we apply the idea to both vertex-streaming and edge-streaming partitioning. This requires defining preassignment rules not only for edges, as in \textsf{2PS}, but also for vertices. Second, instead of using the streaming clustering algorithm of \textsf{2PS}, we compute clusters using CluStRE~\cite{clustre_chhabra_2025}. CluStRE assigns vertices to clusters while streaming using a \emph{modularity} optimizing assignment rule and then applies lightweight restreaming or memetic refinement.
Third, we enforce per-cluster upper bounds on both vertex and edge volume equal to the partition capacity bounds $U_p^\text{vertex}$ and $U_p^\text{vol}$, so that no cluster grows large enough to exceed the capacity of a single partition block. This allows clusters to be mapped to blocks without \hbox{being split.}


To use the clustering for partitioning, we must translate cluster structure into block assignments. Since clustering only groups vertices without assigning them to blocks, we map each cluster to a block, thereby inducing a preferred placement for its vertices. This is necessary because the number of clusters can be significantly larger than the number of blocks, and clusters vary widely in size. Assigning clusters to blocks independently without coordination can therefore lead to severe imbalance, with large clusters overloading individual blocks.

Following the approach of \textsf{2PS}~\cite{mayer20202ps}, we model the cluster-to-block assignment as an instance of makespan scheduling on identical machines~\cite{graham1969makespan}, where blocks correspond to machines, clusters to jobs, and cluster volumes to processing times. The objective is to minimize the maximum total assigned cluster volume across blocks. Since this problem is NP-hard, we employ Graham's classical sorted list scheduling algorithm~\cite{graham1969makespan}. Applied to our setting, clusters are sorted in non-increasing order of volume, and each cluster $C$ is assigned to a block via
\[
  \phi(C) = \arg\min_{p \in \{1,\dots,k\}} \mathit{vol}(p),
\]
where $\mathit{vol}(p)$ denotes the total volume of clusters already assigned to block $p$. This distributes cluster volume across blocks while providing a $4/3$-approximation of the optimal makespan.




The resulting mapping $\phi$ induces a preferred block for each vertex through the block assigned to its cluster. We use this preferred placement to derive preassignments in a second pass over the graph. Unlike \textsf{2PS}, which uses the cluster mapping only for edge prepartitioning, our framework defines preassignment rules for both supported streaming paradigms.

Let $\kappa : V \rightarrow \mathbb{N}$ denote the resulting clustering. For vertex partitioning, a vertex $v$ is preassigned to block $\phi(\kappa(v))$ only if all already preassigned neighbors $u \in N(v)$ satisfy $\phi(\kappa(u)) = \phi(\kappa(v))$, and the preassignment respects balance constraints. Otherwise, $v$ is deferred to the vertex-streaming assignment rule.
For edge partitioning, an edge $(u,v)$ is preassigned to block $\phi(\kappa(u))$ only if $\kappa(u) = \kappa(v)$, and the preassignment respects balance constraints. Otherwise, the edge is deferred to the \hbox{edge-streaming assignment rule.}

Overall, this stage preassigns only those vertices or edges for which the cluster-induced assignment is both locally consistent and feasible under balance constraints. The remaining assignments follow the streaming partitioning rules, exploiting global structure while staying robust to streaming order and capacity constraints.


\subsection{Runtime and Memory Complexity}
\label{subsec:complexity}

\begin{table}
  \small
  \caption{Asymptotic running times. Here, $n=|V|$, $m=|E|$, $k$ is the number of blocks, and $q$ is the number of clusters found in the preprocessing step.}
  \label{tab:streaming_part_runtime_comparison}
  \centering
  \begin{tabular}{lclc}
    \toprule
    \multicolumn{2}{c}{Streaming Vertex Partitioners} & \multicolumn{2}{c}{Streaming Edge Partitioners} \\
    \cmidrule(lr){1-2}\cmidrule(lr){3-4}
    Alg. & Running Time & Alg. & Running Time \\
    \midrule
    \textsf{Random} & $O(n)$
      & \textsf{Random} & $O(m)$ \\
    \textsf{LDG}~\cite{stanton2012streamingLDG} & $O(m{+}nk)$
      & \textsf{HDRF}~\cite{petroni2015hdrf} & $O(mk)$ \\
    \textsf{Fennel}~\cite{tsourakakis2014fennel} & $O(m{+}nk)$
      & 2PS-HDRF~\cite{mayer20202ps} & $O(mk{+}q\log q)$ \\
    \textsf{SIGMA} & $O(m{+}nk)$
      & 2PS-L~\cite{mayer20202ps} & $O(m{+}q\log q)$ \\
    -- & --
      & \textsf{SIGMA} & $O(mk{+}n)$ \\
    \bottomrule
  \end{tabular}
\end{table}


As shown in Table~\ref{tab:streaming_part_runtime_comparison}, both partitioning modes match the dominant asymptotic running times of standard stateful streaming partitioners. For vertex partitioning, the locality term $e(v,p)$ can be computed by aggregating contributions from already assigned neighbors in $O(d(v))$ time, and evaluating the score over all $k$ blocks adds $O(k)$ time per vertex. This yields $O(m+nk)$ total time, matching the asymptotic runtime of \textsf{Fennel}~\cite{tsourakakis2014fennel}. The multi-objective variant preserves the same bound, since the replication estimate is computed from maintained vertex-to-block incidence information.
For edge partitioning, each arriving edge is evaluated against all $k$ blocks using endpoint-affinity and balance terms, yielding $O(k)$ time per edge and $O(mk)$ total time. This matches the asymptotic runtime of stateful edge-streaming partitioners such as \textsf{HDRF}~\cite{petroni2015hdrf}, whose scoring function also evaluates each edge against all blocks.

The clustering phase adds linear work. CluStRE-Light+ processes each vertex together with its neighborhood to evaluate modularity gain over the clusters of already assigned neighbors, and the subsequent cluster-induced preassignment pass scans the graph again. These steps require $O(n+m)$ time for a fixed number of streaming passes. The cluster-to-block mapping sorts clusters by volume. Since cluster volumes are integer values bounded by $2m$, this can be implemented using linear-time integer sorting, e.g., radix sort, in $O(q)$ time. As $q \leq n$, this does not change the preprocessing bound of $O(n+m)$. Consequently, the overall running times are $O(m+nk)$ for vertex partitioning and $O(n+mk)$ for edge partitioning.

The memory footprint is determined by the maintained partition state. The vertex partitioner stores the minimal state required by all stateful streaming partitioners, i.e., vertex-to-block assignment and block load vectors, requiring $O(n+k)$ space. The multi-objective vertex partitioner additionally tracks vertex-to-block incidences to estimate replication, requiring $O(nk)$ space in the worst case. The edge partitioner also maintains vertex-to-block incidences and edge load vectors, and therefore also requires $O(nk)$ worst-case memory. Clustering information, cluster volumes, degrees, cluster-to-block mapping and the auxiliary array used for radix sorting require $O(n+q+k)$ additional space with $q \leq n$, and do not change the overall asymptotic memory bounds.

\section{Experimental Setup}

\subsection{Datasets}
\label{sec:4.1_datasets}
We evaluate our approach on a diverse collection of widely used benchmark graph-datasets, including \textit{Amazon Computers, Flickr, Reddit, Twitch, ogbn-arxiv}, and \textit{ogbn-products} as shown in Table~\ref{tab:datasets}. These datasets span multiple domains and vary significantly in terms of scale, graph density, and feature characteristics. This diversity ensures that our evaluation captures both scalability and robustness across different graph types.
Before partitioning, we symmetrize each graph, merge duplicate edges, and remove self-loops while retaining all vertices. Table~\ref{tab:datasets} reports the structural statistics of the resulting simple undirected graphs.

\subsection{GNN Systems} We select two different distributed GNN systems that support different training and graph partitioning strategies, following the benchmarking study of Merkel et al.~\cite{merkel2023experimental}. DistGNN~\cite{md2021distgnn} supports edge partitioning and employs distributed full-graph training. DistDGL~\cite{zheng2020distdgl} supports vertex partitioning and employs distributed mini-batch training using neighbor-based sampling. 
Note: Due to scalability limitations of DistGNN, we could not obtain results for \textit{ogbn-products} with this system.

\begin{table*}[t]
  \centering
  \small
  \caption{Overview and structural characteristics of the graph datasets.
  Undirected edges are counted once (twice in bidirected storage). 
  Mean degree is computed as $2|E|/|V|$ from unrounded counts, diameter on the largest connected component, and $C$ denotes the average clustering coefficient.}
  \label{tab:datasets}
  \resizebox{\textwidth}{!}{%
    \begin{tabular}{llrrrrrrrr}
      \toprule
      Graph & Type
      & \#Vertices
      & \#Edges
      & Diameter
      & $C$
      & Mean degree
      & Median degree
      & Degree std.
      & Max degree \\
      \midrule
      \textit{Amazon Computers}~\cite{shchur2018pitfalls}
      & e-commerce
      & 13.8k
      & 245.9k
      & 10
      & 0.344126
      & 35.7564 
      & 22
      & 70.3056 
      & 2992 \\

      \textit{Flickr}~\cite{zeng2019graphsaint}
      & social
      & 89.2k
      & 449.9k
      & 8
      & 0.032957
      & 10.0813 
      & 6
      & 31.7488 
      & 5425 \\

      \textit{Twitch}~\cite{rozemberczki2021twitch}
      & social
      & 168.1k
      & 6.8M
      & 8
      & 0.159925 
      & 80.8684 
      & 32
      & 314.1618 
      & 35279 \\

      \textit{ogbn-arxiv}~\cite{hu2020opengraphbenchmark}
      & citation/papers
      & 169.3k
      & 1.2M
      & 25
      & 0.226129 
      & 13.674
      & 6
      & 68.6381 
      & 13161 \\

      \textit{Reddit}~\cite{hamilton2017inductive}
      & social
      & 233.0k
      & 57.3M
      & 8
      & 0.579322
      & 491.9876 
      & 230
      & 799.0102 
      & 21657 \\

      \textit{ogbn-products}~\cite{hu2020opengraphbenchmark}
      & product/co-purchase
      & 2.4M
      & 61.9M
      & 29
      & 0.410788 
      & 50.5172
      & 26
      & 95.9050 
      & 17481 \\

      \bottomrule
    \end{tabular}%
  }
\end{table*}

\subsection{Baseline Partitioning Algorithms}
\label{sec:4.2_baselines}
To comprehensively assess the effectiveness of \textsf{SIGMA}, we compare against a broad set of established partitioning algorithms, covering both edge-based and vertex-based strategies.

For edge partitioning, we consider seven state-of-the-art partitioners from different categories: \textsf{HDRF}~\cite{petroni2015hdrf} for stateful streaming, \textsf{Random Partitioning} and \textsf{DBH}~\cite{xie2014distributedDBH} for stateless streaming, \textsf{HeiStreamE}~\cite{chhabra2024heiStreamE} for buffered streaming, \textsf{2PS}~\cite{mayer20202ps} for multi-pass streaming, \textsf{FSM}~\cite{liu2024fsm} and \textsf{HEP}~\cite{mayer2021hybridHEP} for in-memory partitioning. \textsf{HEP} introduces a parameter $\tau$ that controls the ratio between streaming and in-memory partitioning; we use $\tau = 100$, which comes close to full in-memory partitioning.

For vertex partitioning, we evaluate against ten state-of-the-art partitioners from different categories: \textsf{METIS}~\cite{karypis1997metis}, \textsf{KaHIP}~\cite{sanders2013kahip} and \textsf{PuLP}~\cite{slota2014pulp} for in-memory partitioning, 
\textsf{Random Partitioning} (integrated with DGL) for stateless streaming, 
\textsf{LDG}~\cite{stanton2012streamingLDG} and 
\textsf{FENNEL}~\cite{tsourakakis2014fennel} for stateful streaming, \textsf{SPRING} (\textsf{SDT-GNN})~\cite{huang2024catgnn} for clustering-based streaming, and \textsf{Cuttana}~\cite{hajidehi2023cuttana}, \textsf{HeiStream}~\cite{faraj2022bufferedHeiStream}, and \textsf{BuffCut}~\cite{baumgairtner2026buffcut} for buffered streaming.

\textsf{SIGMA} is employed in both edge and vertex partitioning settings, due to its ability to perform both paradigms.
For all baseline partitioners, we used the default configurations provided by their implementations or settings documented in their original publications, without additional hyperparameter tuning. Exact configurations are provided in the supplementary material.

\subsection{System Configuration}
All experiments are conducted on a local compute cluster comprising four physical servers. For $k=2$ and $k=4$ partitions, we use two and four servers, respectively. For $k=8$, we distribute the partitions evenly across all four servers.
Each server is equipped with an AMD EPYC 9454P processor (48 cores), 768 GB DDR5 ECC memory, two NVIDIA L40S GPUs with 48 GB GDDR6 memory each, and SATA datacenter SSD storage. 


\subsection{GNN Training Configuration}
We train a two-layer GraphSAGE model~\cite{hamilton2017inductive}. The model is implemented with SAGEConv layers using the GCN aggregator and a hidden dimension of 16 to enable fast training even on the largest graphs.
Each hidden layer is followed by a ReLU activation function and dropout with probability 0.5. Training is performed using the Adam optimizer with a learning rate of 0.003 and weight decay of $5 \times 10^{-4}$; each experiment is run for 100 epochs across three independent runs to report mean performance and standard deviation. These settings are representative of realistic GNN workloads.

The choice of GNN architecture is motivated by its broad applicability and availability in GNN training systems.
To ensure comparability, we keep the model architecture, optimization procedure, and hyperparameters consistent across all partitioning methods. This allows us to isolate the impact of the partitioning strategy on training efficiency and model performance. 
To assess the impact of the GNN architecture, we additionally evaluate a Graph Attention Network (GAT) \cite{velivckovic2017graph-GAT} on \textit{Amazon Computers} for $k \in \{2,4,8\}$ using both DistGNN and DistDGL. For both frameworks, we set \texttt{n\_layers}=1, \texttt{n\_hidden}=8 and use a learning rate of 0.003.

For DistDGL, mini-batch training uses a standard batch size of 1,000 and neighbor sampling fanouts of [25,25] for the two GraphSAGE layers. As mentioned above, DistGNN employs full-graph training, so there are no batch sizes and fanouts to be set.

\subsection{Evaluation Metrics}
We evaluate partitioning strategies from a holistic perspective, focusing on their impact on communication cost, load balance, and end-to-end training efficiency in distributed GNN execution. By combining these metrics, we provide a comprehensive evaluation of how partitioning strategies affect both the scalability and effectiveness of distributed GNN training.

\emph{Cut size.}
For vertex partitioning, we measure the normalized edge cut $\lambda$ based on $E_{\mathrm{cut}}$. For edge partitioning, we measure replication factor. See Section~\ref{sec:2_background} for detailed specifications.


\emph{Balance.}
We report balance metrics corresponding to the constraints defined in Section~\ref{sec:2_background}. In particular, we measure vertex and edge balance based on block sizes. For vertex partitioning, we report vertex balance and edge balance over $V_i$ and $E_i$, respectively. For edge partitioning, we report edge balance over $E_i$ and vertex balance over $V(E_i)$.

\emph{Runtime.}
We evaluate system efficiency using \emph{partitioning time} and average \emph{training time} per epoch. Partitioning time captures the preprocessing overhead required to generate graph partitions, while per-epoch training time reflects the steady-state computational cost during distributed GNN training.
For each framework and model, we rank methods by mean epoch time within each dataset–partition-count configuration and average ranks equally across configurations.

\emph{Memory Usage.}
We measure peak host memory consumption using the Resident Set Size (RSS) per worker, which represents the maximum amount of physical main memory occupied by each worker process during training.



\section{Results and Evaluation}
This section evaluates the proposed partitioning approach from two perspectives: (i) structural partition quality and (ii) its impact on distributed GNN training performance. 

\subsection{Sensitivity Analysis of $\gamma$ and $\tau$}
We first validate the default parameters $\gamma = 2.5$ and $\tau = 0.5$ proposed in Section~\ref{subsec:vertex_partitioning} via a full-factorial sensitivity study. We sweep $\gamma \in \{1.5, 2.0, 2.5, 3.0, 3.5\}$ and $\tau \in \{0.125, 0.25, 0.5, 0.75, 1.0\}$ at each of $k \in \{2, 4, 8, 16, 32\}$. Each of the 25 $(\gamma,\tau)$ configurations is evaluated on five held-out web and social graphs (\textit{in-2004} and \textit{web-Google}~\cite{10.1145/1963405.1963488, boldi2004webgraph, 10.1145/3160017}, and \textit{wiki-Talk}, \textit{coAuthorsDBLP}, and \textit{com-amazon}~\cite{snapnets}) distinct from the six benchmark graphs (Table ~\ref{tab:datasets}) used in the partitioning quality and system performance evaluations. We average edge-cut ratio, replication factor, vertex balance, and edge balance over $k$ and graphs, min--max normalize each across the 25 configurations, and use their mean as global score. A score $s$ thus lies on average a fraction $s$ of the way from the best to the worst configuration of this grid.

Table~\ref{tab:sensitivity} reports the top candidates. The default $(2.5,0.5)$ ties with $(2.5,0.25)$ for the best global score (difference below $10^{-5}$) and is retained for its lower replication factor the objective $\tau$ targets. For every $\gamma$, the best $\tau$ is 0.25 or 0.5. Lowering $\tau$ to 0.125 yields the highest replication factor, whereas raising it to 1.0 reduces replication no further but increases the edge-cut ratio by up to 2.8\%. The effect of $\gamma$ is flat except at $\gamma = 1.5$, which underperforms for nearly every $\tau$. Since raw averages span only 0.259--0.270 in edge-cut ratio and 1.858--1.880 in replication factor across the grid, and the top five differ by at most 1.1\% on any metric, \textsf{SIGMA} is robust to the exact choice of $\gamma$ and $\tau$ around the default.

\begin{table}[t]
\caption{Top-5 $(\gamma,\tau)$ configurations by normalized global score (0 = best, 1 = worst of 25; lower is better) and raw metric averages over five graphs and $k \in \{2,\dots,32\}$. Bottom rows vary the default by low $\tau$, high $\tau$, and low $\gamma$. Best per column among listed rows in bold.}
\label{tab:sensitivity}
\small
\begin{tabular}{cc|c|cccc}
\toprule
$\gamma$ & $\tau$ & global & edge-cut & repl. & vertex bal. & edge bal. \\
\midrule
2.5 & 0.5 & \textbf{0.102} & 0.2593 & \textbf{1.8589} & 1.0252 & 1.0750 \\
2.5 & 0.25 & \textbf{0.102} & \textbf{0.2586} & 1.8665 & \textbf{1.0235} & \textbf{1.0741} \\
2.0 & 0.25  & 0.143 & 0.2592 & 1.8642 & 1.0244 & 1.0752 \\
3.0 & 0.5 & 0.158 & 0.2609 & 1.8609 & 1.0252 & 1.0749 \\
3.5 & 0.5 & 0.185 & 0.2615 & 1.8636 & 1.0249 & 1.0745 \\
\midrule
2.5 & 1.0 & 0.481 & 0.2665 & \textbf{1.8589} & 1.0292 & 1.0795 \\
2.5 & 0.125 & 0.559 & 0.2611 & 1.8796 & 1.0237 & 1.0848 \\
1.5 & 0.5 & 0.682 & 0.2677 & 1.8699 & 1.0300 & 1.0808 \\
\bottomrule
\end{tabular}
\end{table}

\subsection{Partitioning Quality}
\label{sec:5.1_partitioning-quality}
\begin{figure*}[h]
    \centering
    \includegraphics[width=\textwidth]{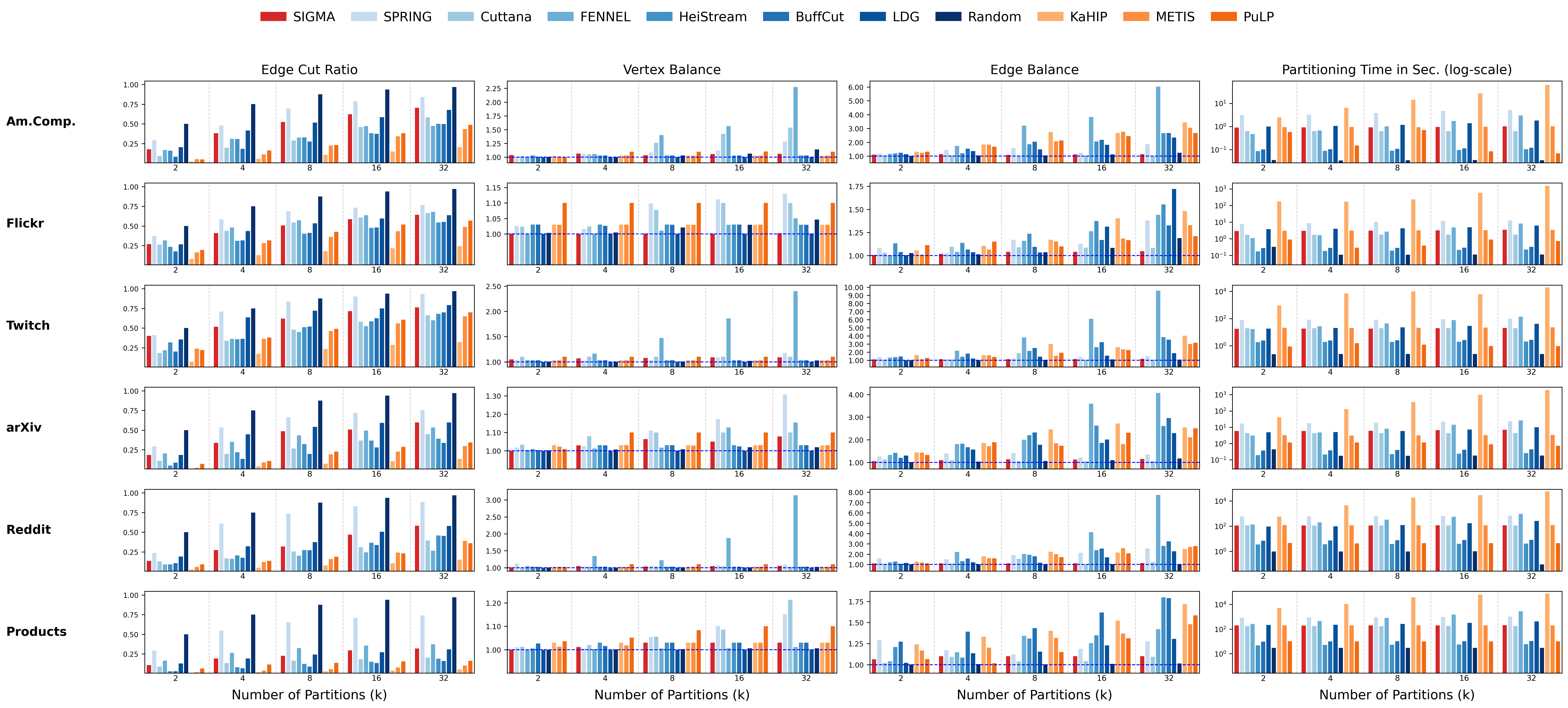}
    \caption{Quality metrics for vertex partitioners by dataset and partitioner.}
    \label{fig:combined_quality_metrics-NodePartitioner}
\end{figure*}

We evaluate partition quality across a range of $k$-values (number of partitions). We report vertex and edge balance along with the corresponding cut size metric (replication factor for edge partitioning, edge-cut ratio for vertex partitioning). Additionally, we report the partitioning time. We emphasize that different from existing work on vertex and edge partitioning, we report \underline{both} balancing metrics for both partitioning paradigms. 
In Figures \ref{fig:combined_quality_metrics-EdgePartitioner}-\ref{fig:combined_memory_metrics-EdgePartitioner}, blue and orange bars represent streaming and in-memory partitioning approaches, respectively, while dashed lines indicate (theoretical) optimal values where applicable.

\subsubsection{Edge Partitioning}
\begin{figure*}[h]
    \centering
    \includegraphics[width=\textwidth]{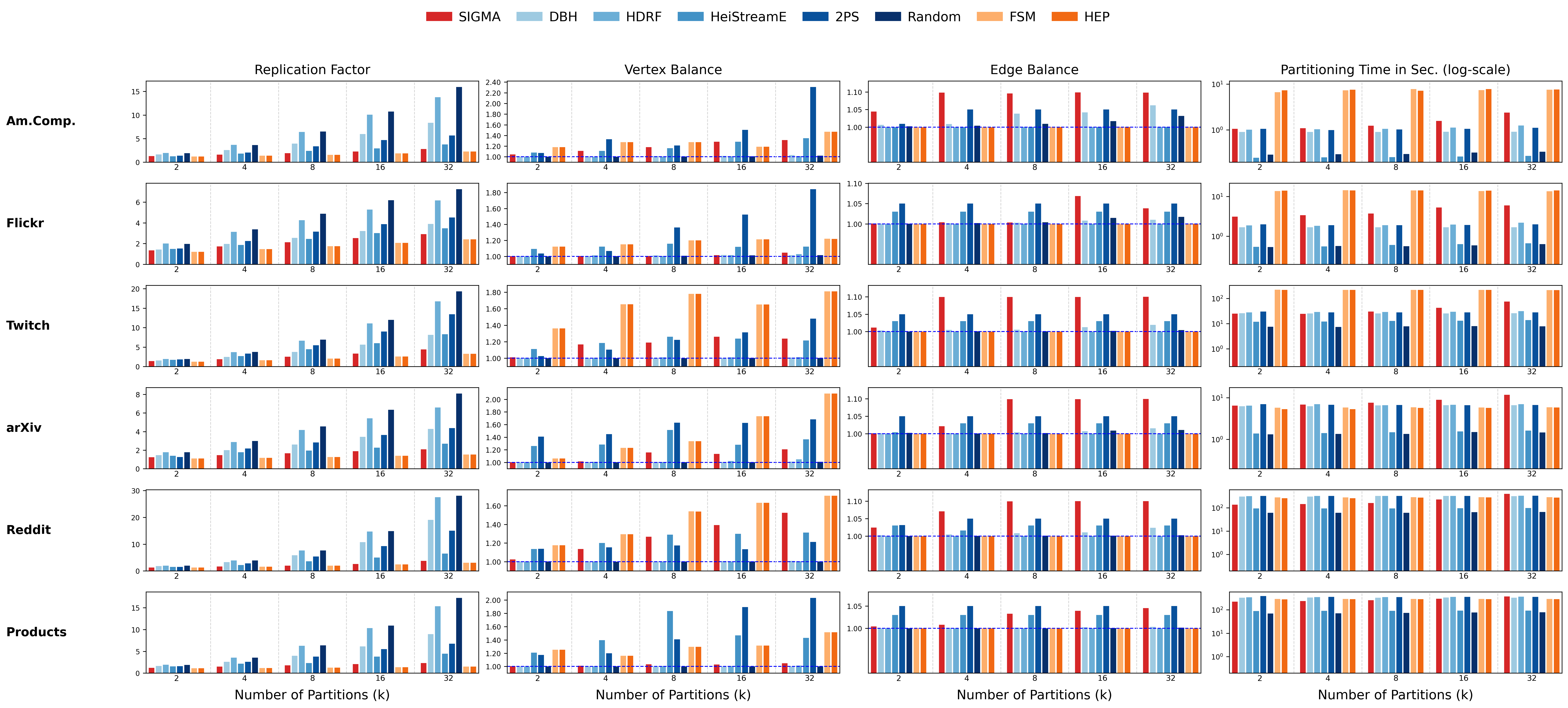}
    \caption{Quality Metrics for edge partitioners by dataset and partitioner.}
    \label{fig:combined_quality_metrics-EdgePartitioner}
\end{figure*}

The complete results are shown in  Figure~\ref{fig:combined_quality_metrics-EdgePartitioner}. We discuss the results metric by metric.

\paragraph{Replication factor.} 
\textsf{SIGMA} achieves the lowest replication factor among all streaming partitioners across the evaluated datasets and partition counts, with its advantage increasing for larger $k$. On \textit{Amazon Computers} at $k=32$, \textsf{SIGMA} attains a replication factor of $2.80$, compared to $3.74$ for \textsf{HeiStreamE}, $13.8$ for \textsf{HDRF}, and $16.0$ for \textsf{Random}. Similar improvements are observed on larger graphs: on \textit{Reddit} and \textit{ogbn-products}, \textsf{SIGMA} reduces replication by approximately 87\% and 81\% compared to \textsf{Random}, respectively, and by approximately 82\% compared to \textsf{HDRF} on both datasets at $k=32$. The hybrid in-memory approaches \textsf{FSM} and \textsf{HEP} achieve the lowest absolute replication factors, although \textsf{SIGMA} remains competitive.

\paragraph{Balancing.}
(1) Edge Balance:
Unlike some competing partitioners that enforce strict balance, \textsf{SIGMA} supports a configurable edge imbalance factor. This additional flexibility allows the partitioning process to explore a larger solution space, potentially improving partition quality for GNN training. In all experiments, we permitted \textsf{SIGMA} an imbalance factor of $1.10$ and consistently satisfied this constraint.

(2) Vertex Balance:
\textsf{SIGMA} achieves lower vertex imbalance than the in-memory partitioners across all evaluated datasets and partition counts (with one single exception, \textit{Amazon Computers} at $k=16$). Specifically, \textsf{SIGMA}'s vertex balance ranges from $1.00$ to $1.53$, whereas \textsf{HEP} and \textsf{FSM} exhibit values between $1.06$ and $2.09$. 
The maximum observed balance factor is reduced from 2.09 to 1.53, corresponding to an approximately 51\% reduction in imbalance relative to the ideal balance factor of $1.0$. 
Although some streaming partitioners, such as \textsf{Random}, \textsf{DBH}, and \textsf{HDRF}, achieve lower vertex imbalance than \textsf{SIGMA}, \textsf{SIGMA} consistently outperforms the high-quality partitioners \textsf{HeiStreamE} and \textsf{2PS}, particularly on larger graphs and at higher partition counts. For example, on \textit{ogbn-products} with $k=32$, \textsf{SIGMA} attains a vertex balance of $1.05$, compared to $2.03$ for \textsf{2PS} and $1.43$ for \textsf{HeiStreamE}. Relative to the ideal balance factor of $1.0$, this corresponds to reductions in vertex imbalance of approximately 95\% and 88\%, respectively.

\paragraph{Partitioning time.}
\textsf{SIGMA} incurs moderate overhead compared to faster streaming approaches such as \textsf{HeiStreamE}, but remains competitive with \textsf{HDRF}, \textsf{DBH}, and \textsf{2PS}, particularly on larger graphs. It also frequently outperforms in-memory approaches in partitioning time, especially on smaller datasets.


\subsubsection{Vertex Partitioning}
The complete results are shown in  Figure~\ref{fig:combined_quality_metrics-NodePartitioner}. We discuss the results metric by metric.

\paragraph{Edge-cut ratio.} 
\textsf{SIGMA} ranks among the mid-performing approaches in terms of edge-cut ratio, due to its stronger focus on multi-objective balancing.
It consistently outperforms \textsf{Random} and \textsf{SPRING} across all evaluated configurations. However, in-memory partitioners such as \textsf{METIS}, \textsf{KaHIP}, and \textsf{PuLP} generally achieve lower edge cuts, while \textsf{FENNEL} and buffered streaming approaches also outperform \textsf{SIGMA} in some settings. For instance, on \textit{Amazon Computers} at $k=32$, \textsf{FENNEL} achieves $0.475$ compared to \textsf{SIGMA} achieving $0.704$. These differences reflect \textsf{SIGMA}'s additional emphasis on vertex replication and multi-dimensional balance rather than edge-cut minimization alone.

\paragraph{Balancing.} (1) Vertex Balance:
On vertex balance, \textsf{SIGMA} remains consistently close to the optimum value of $1.0$ across all datasets and partition counts, with values ranging from $1.00$ to $1.09$. Compared to \textsf{FENNEL}, which suffers of a vertex balance factor of $2.40$ on \textit{Flickr} for $k=32$, \textsf{SIGMA} reduces imbalance by approximately 94\%.

(2) Edge Balance:
A similar trend can be observed for edge balance. Across all datasets and partition counts, \textsf{SIGMA} maintains balance factors between $1.01$ and $1.18$, remaining close to the ideal value of $1.0$. On \textit{Twitch} and \textit{ogbn-arxiv}, only \textsf{Random} achieves slightly better edge balance than \textsf{SIGMA} for $k>4$. On \textit{Flickr}, only \textsf{Random} consistently outperforms \textsf{SIGMA} for $k>4$, with \textsf{LDG} achieving a marginally better result only at $k=8$. Overall, \textsf{SIGMA} consistently ranks first or among the top-performing approaches with respect to both vertex and edge balance, demonstrating that its ability to achieve highly balanced partitioning across both balancing dimensions.

\paragraph{Partitioning time.}
\textsf{SIGMA} remains competitive with streaming approaches such as \textsf{Cuttana}, \textsf{FENNEL}, and \textsf{LDG}, typically differing by only a few seconds. It is, however, consistently outperformed by \textsf{HeiStream}, \textsf{Random}, and \textsf{BuffCut}. Among the in-memory approaches, \textsf{METIS} and \textsf{PuLP} achieve runtimes comparable to the streaming partitioners, whereas \textsf{KaHIP} exhibits by far the highest partitioning times across all evaluated datasets and partition counts.

\subsubsection{Discussion}
Overall, the results demonstrate that \textsf{SIGMA} achieves a favorable trade-off between partition quality, balance, and partitioning cost. For edge partitioning, \textsf{SIGMA} consistently attains the lowest replication factors among all streaming approaches while maintaining good edge and vertex balance. For vertex partitioning, \textsf{SIGMA} achieves  competitive edge-cut ratios and consistently ranks among the best methods in terms of vertex and edge balance. Although individual partitioners occasionally outperform \textsf{SIGMA} on specific metrics, these advantages typically come at the cost of performing worse according to another criterion.
In contrast, \textsf{SIGMA} delivers robust performance across all considered metrics and datasets, without exhibiting major weaknesses in any single category. These results indicate that \textsf{SIGMA} successfully balances the conflicting objectives of minimizing communication overhead, maintaining balanced partitions with regard to vertices and edges, and preserving the scalability characteristics required for streaming graph partitioning.


\subsection{Impact on Distributed GNN Training}
We evaluate how partition quality affects GNN training time and memory consumption.

\subsubsection{Training performance}
We report mean training time per epoch. Combined with the partitioning time reported in Section~\ref{sec:5.1_partitioning-quality}, these measurements allow all relevant runtime comparisons, including total training time and end-to-end runtime comprising partitioning and GNN training.

\begin{figure*}[h]
    \centering
    \includegraphics[width=0.97\textwidth]{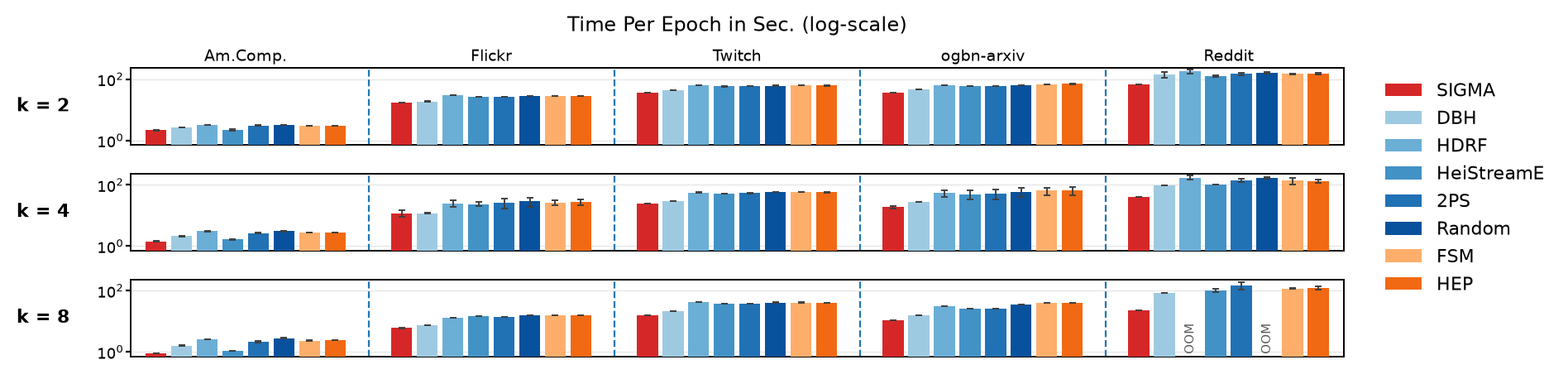}
    \caption{Mean time per epoch of GNN training under edge partitioning (DistGNN) by dataset and partitioner.}
    \label{fig:combined_time_metrics-EdgePartitioner}
\end{figure*}

\paragraph{Edge Partitioning (DistGNN)}
For edge-partitioning based training on DistGNN, 
Figure~\ref{fig:combined_time_metrics-EdgePartitioner} shows that \textsf{SIGMA} achieves the lowest mean epoch time in all 15 evaluated dataset–partition-count configurations, corresponding to a mean rank of 1.00 across all methods.
At equal partition counts, \textsf{SIGMA} reduces epoch time on \textit{Twitch} from approximately 15\% relative to \textsf{DBH} at $k=2$ to 63\% relative to \textsf{HDRF} at $k=8$. On \textit{Amazon Computers}, epoch time reductions range from approximately 1\% relative to \textsf{HeiStreamE} at $k=2$ to 67\% relative to \textsf{Random} at $k=8$.

\textsf{SIGMA} also consistently outperforms competitors on \textit{Flickr}, \textit{ogbn-arxiv}, and \textit{Reddit}, reducing epoch time by approximately 2--62\%, 23--73\%, and 47--84\%, respectively, depending on the baseline and partition count. Against the in-memory partitioners \textsf{FSM} and \textsf{HEP}, reductions are approximately 38--62\% on \textit{Flickr}, 48--73\% on \textit{ogbn-arxiv}, and 54--81\% on \textit{Reddit}.

On \textit{Reddit} at $k=8$, \textsf{HDRF} and \textsf{Random} run out of memory (OOM) as GNN training exhausts all available memory of our cluster.

\begin{figure*}[h]
    \centering
    \includegraphics[width=\textwidth]{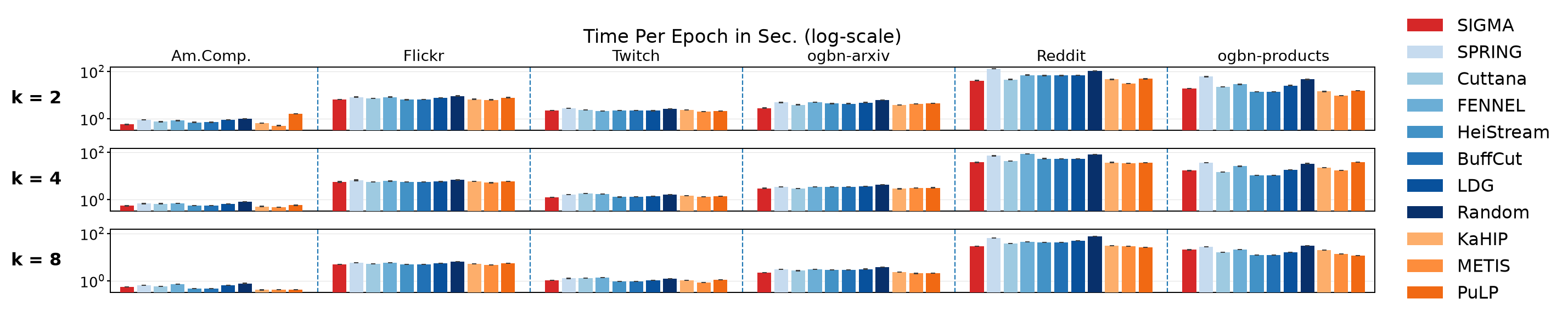}
    \caption{Mean time per epoch of GNN training under vertex partitioning (DistDGL) by dataset and partitioner.}
    \label{fig:combined_time_metrics-NodePartitioner}
\end{figure*}

\paragraph{Vertex Partitioning (DistDGL)}
Figure~\ref{fig:combined_time_metrics-NodePartitioner} shows that \textsf{SIGMA} achieves competitive mean epoch times for vertex-partitioning based training on DistDGL, with relative performance varying by dataset and partition count. \textsf{SIGMA} outperforms \textsf{FENNEL} and \textsf{LDG} at all evaluated partition counts on \textit{Amazon Computers}, \textit{Flickr}, \textit{ogbn-arxiv}, and \textit{Reddit}, 
while on \textit{Twitch} and \textit{ogbn-products}, \textsf{SIGMA} outperforms \textsf{FENNEL} in four of the six configurations and \textsf{LDG} in three.
It achieves the lowest epoch time among all evaluated methods on \textit{ogbn-arxiv} at $k=2$ and \textit{Twitch} at $k=4$. On \textit{ogbn-arxiv} at $k=2$, its epoch time is approximately 25\% lower than that of the fastest competitor, \textsf{KaHIP}. On \textit{Reddit}, \textsf{SIGMA} outperforms all evaluated streaming baselines at every partition count, reducing epoch time by approximately 9--69\%, depending on the baseline and partition count.

On \textit{ogbn-products}, \textsf{SIGMA} trails \textsf{HeiStream} and \textsf{BuffCut} at all evaluated partition counts, as well as all in-memory methods at $k=2$ and $k=8$, but outperforms \textsf{KaHIP}, \textsf{METIS}, and \textsf{PuLP} at $k=4$. 
Across the 18 configurations, \textsf{SIGMA} achieves the best mean rank among streaming methods (2.22) and the second-best mean rank overall (3.67), behind \textsf{METIS} (2.11).


\subsubsection{Memory Consumption}
We next examine how partitioning characteristics affect per-worker memory consumption during training. 

\begin{figure*}[h]
    \centering
    \includegraphics[width=0.97\textwidth]{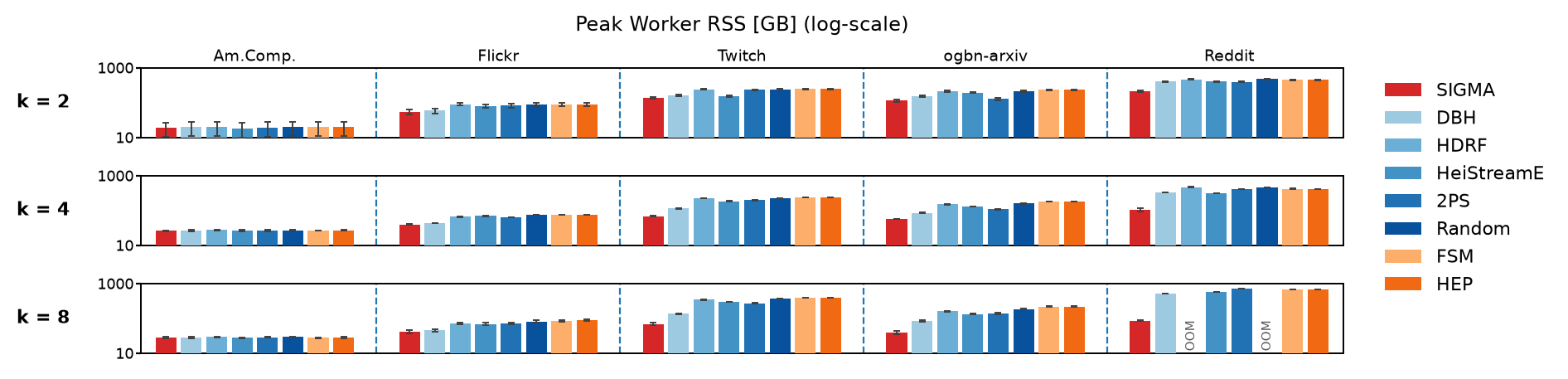}
    \caption{Memory consumption of GNN training under edge partitioning (DistGNN): peak RAM usage across workers.}
    \label{fig:combined_memory_metrics-EdgePartitioner}
\end{figure*}

\paragraph{Edge Partitioning (DistGNN)}
Figure~\ref{fig:combined_memory_metrics-EdgePartitioner} shows that \textsf{SIGMA} achieves the lowest peak RAM usage on \textit{Flickr}, \textit{Twitch}, \textit{ogbn-arxiv}, and \textit{Reddit} at all evaluated partition counts ($k \in \{2,4,8\}$), while memory consumption on \textit{Amazon Computers} remains similar across partitioners. \textsf{SIGMA}'s advantage is most pronounced on larger datasets and at higher partition counts. At $k=8$, \textsf{SIGMA} uses approximately $70$ GB on \textit{Twitch} and $40$ GB on \textit{ogbn-arxiv}, substantially less than all competitors. The largest difference occurs on \textit{Reddit} at $k=8$, where \textsf{SIGMA} uses approximately $90$ GB compared with over $500$ GB for all other methods completing training, while \textsf{HDRF} and \textsf{Random} run out of memory.

\paragraph{Vertex Partitioning (DistDGL)}

We measured both host and GPU memory usage for DistDGL but omit
a detailed comparison here because neither provides an unambiguous
indicator of partitioning quality in this setting. Host memory includes partition structures, locally owned features, and sampling and system-level overhead; ghost-node features are not persistently replicated in each partition~\cite{doku-dgl_partition_2024}. GPUs process sampled computation blocks and their required features, so device memory primarily reflects the sampled mini-batch rather than the complete partition. Consequently, lower peak memory usage does not necessarily imply better partitioning quality or faster training. Detailed memory measurements are provided in the supplementary material.

\subsubsection{GAT Training}
To assess whether the observed performance generalizes beyond GraphSAGE, we additionally evaluate GAT training on \textit{Amazon Computers} for $k \in \{2,4,8\}$. Under DistGNN, \textsf{SIGMA} achieves the lowest epoch times and peak RAM usage across all evaluated partitioners and values of $k$, absolute differences are small. 
Under DistDGL, \textsf{SIGMA} ranks first among streaming methods at $k=2$ and $k=4$, and third at $k=8$, behind \textsf{HeiStream} and \textsf{BuffCut}.
Across the three partition counts, \textsf{SIGMA} achieves mean epoch-time ranks of 1.00 under DistGNN and 4.00 under DistDGL across all methods, and the best mean rank among streaming methods in both frameworks.
Overall, these results demonstrate that \textsf{SIGMA} remains competitive across GNN architectures, particularly with edge partitioning. Detailed results are provided in the supplementary material.

\vspace{0.5cm}
\textbf{Summary and Discussion.}
Overall, \textsf{SIGMA} achieves a favorable balance between partitioning overhead and training efficiency. 

These results can be explained by the partition-quality characteristics observed in Section~\ref{sec:5.1_partitioning-quality}. For edge partitioning, \textsf{SIGMA} consistently achieves substantially lower replication factors while maintaining good edge and vertex balance. Lower replication reduces duplicated vertex information that must be stored, communicated, and synchronized across workers, thereby reducing communication overhead and memory footprint while improving training throughput. Strong edge and vertex balance also prevents individual workers from becoming bottlenecks, ensuring efficient hardware utilization. For vertex partitioning, \textsf{SIGMA} combines competitive edge-cut ratios with excellent balance, reducing inter-partition communication while evenly distributing computation and state. Thus, improved partition quality directly translates into shorter training times and lower memory consumption for edge-partitioned training.

\section{Related Work}
Graph partitioning traditionally targets edge-cut minimization through vertex partitioning or vertex-replication minimization through edge partitioning. In-memory partitioners such as \textsf{METIS}~\cite{karypis1997metis}, \textsf{KaHIP}~\cite{sanders2013kahip}, and \textsf{KaMinPar}~\cite{DeepMultilevelGraphPartitioning} employ multilevel techniques to produce high-quality balanced partitions. Extensions include multi-constraint partitioning~\cite{Karypis1998MultiConstraint}, which balances multiple vertex weights, and multi-objective approaches such as \textsf{PuLP}~\cite{slota2014pulp}, which optimize several criteria (e.g., edge cut and replication factor). However, these methods require random access to the graph and substantial memory, limiting their scalability compared to streaming approaches.

Streaming partitioners instead assign vertices or edges incrementally using limited state. One-pass approaches include \textsf{LDG}~\cite{stanton2012streamingLDG} and \textsf{Fennel}~\cite{tsourakakis2014fennel} for vertex partitioning, and \textsf{DBH}~\cite{xie2014distributedDBH}, \textsf{Grid} and \textsf{PDS}~\cite{Jain2013}, and \textsf{HDRF}~\cite{petroni2015hdrf} for edge partitioning. Restreaming methods, such as \textsf{ReLDG} and \textsf{ReFennel}~\cite{Nishimura2013} and \textsf{2PS}~\cite{mayer20202ps}, improve partition quality through additional passes. Buffered approaches exploit partial structural information by temporarily processing batches, including \textsf{HeiStream}~\cite{faraj2022bufferedHeiStream}, \textsf{Cuttana}~\cite{hajidehi2023cuttana}, \textsf{BuffCut}~\cite{baumgairtner2026buffcut}, and \textsf{LocalDGP}~\cite{ji2025localdgp} for vertex partitioning, and \textsf{HeiStreamE}~\cite{chhabra2024heiStreamE}, \textsf{HoVerCut}~\cite{Sajjad2016}, and \textsf{ADWISE}~\cite{Mayer2018Adwise} for edge partitioning. Hybrid methods such as \textsf{FSM}~\cite{liu2024fsm} and \textsf{HEP}~\cite{mayer2021hybridHEP} combine streaming and in-memory components, maintaining additional global state to guide assignments. \textsf{FSM} additionally balances both vertices and edges but relies on in-memory state and supports only edge partitioning. To the best of our knowledge, no existing streaming framework combines multiple optimization objectives with hard balance constraints across several load dimensions while supporting both vertex and edge partitioning. For broader overviews, we refer to existing surveys~\cite{Buluc2016, ccatalyurek2023more, Schulz2019survey}.

Prior work has also addressed graph partitioning for distributed GNN training. GLISP~\cite{zhu2024glisp} and LPS-GNN~\cite{cheng2025lps} consider partitioning within broader training and deployment frameworks. ARMADA~\cite{waleffe2025armada} introduces GREM, a min-edge-cut objective for vertex partitioning that improves streaming greedy assignments by revisiting earlier decisions, while CATGNN~\cite{huang2024catgnn} integrates the streaming vertex partitioner \textsf{SPRING}. Unlike these vertex-oriented approaches, \textsf{SIGMA} focuses on the partitioning problem itself and supports both vertex and edge partitioning within a unified framework.

\section{Conclusion}
We presented \textsf{SIGMA}, a unified streaming framework for both vertex and edge graph partitioning. Motivated by the observation that distributed GNN training simultaneously depends on communication efficiency and workload balance of both vertices and edges, \textsf{SIGMA} jointly accounts for these objectives while retaining the scalability advantages of streaming partitioning. A clustering-based preprocessing stage further improves partition quality by incorporating global graph structure.

Our experimental evaluation on six benchmark graphs and two distributed GNN training systems demonstrates that \textsf{SIGMA} consistently achieves favorable trade-offs between partition quality, training efficiency, and memory consumption.


Beyond the specific algorithmic improvements, our results show that vertex and edge partitioning need not be treated as entirely separate problems. A unified streaming framework can effectively support both paradigms, reducing engineering complexity and enabling shared advances across a broad range of distributed graph processing.



\bibliographystyle{ACM-Reference-Format}
\bibliography{sample-base}


\end{document}